\documentclass[twocolumn,superscriptaddress,letter]{revtex4}%
\usepackage{amssymb}
\usepackage{color}
\usepackage{graphicx}
\usepackage{dcolumn}
\usepackage{bm}
\usepackage[header,title,page,titletoc]{appendix}
\usepackage{amsmath}
\usepackage{amsfonts}%
\providecommand{\U}[1]{\protect\rule{.1in}{.1in}}

\begin{document}
\title{Effective non-Hermitian Physics for Degenerate Ground States of non-Hermitian
Ising Model with $\mathcal{RT}$-Symmetry}
\author{Can Wang}
\affiliation{Center for Advanced Quantum Studies, Department of Physics, Beijing Normal
University, Beijing 100875, China}
\author{Meng-Lei Yang}
\affiliation{Center for Advanced Quantum Studies, Department of Physics, Beijing Normal
University, Beijing 100875, China}
\author{Cui-Xian Guo}
\affiliation{Center for Advanced Quantum Studies, Department of Physics, Beijing Normal
University, Beijing 100875, China}
\author{Xiao-Ming Zhao}
\affiliation{Beijing National Laboratory for Condensed Matter Physics, Institute of
Physics, Chinese Academy of Sciences, Beijing 100190, China}
\author{Su-Peng Kou}
\thanks{Corresponding author}
\email{spkou@bnu.edu.cn}
\affiliation{Center for Advanced Quantum Studies, Department of Physics, Beijing Normal
University, Beijing 100875, China}

\begin{abstract}
In this paper, based on a one dimensional non-Hermitian spin model with
$\mathcal{RT}$-invariant term, we study the non-Hermitian physics for the two
(nearly) degenerate ground states. By using the high order perturbation
method, an effective pseudo-spin model is obtained to describe non-Hermitian
physics for the two (nearly) degenerate ground states, which are precisely
consistent with the numerical calculations. We found that there may exist
effective (anti) $\mathcal{PT}$-symmetry for the effective pseudo-spin model
of the two (nearly) degenerate ground states. In particular, there exists
spontaneous (anti) $\mathcal{PT}$-symmetry breaking for the topological
degenerate ground states with tunable parameters in external fields. We also
found that even a very tiny imaginary external field applied will drive
$\mathcal{PT}$ phase transition.

\end{abstract}

\maketitle

\section{Introduction}

For a class of non-Hermitian quantum systems as dynamic equilibrium steady
systems under specific symmetry, their Hamiltonians may exhibit entirely real
eigenvalues like Hermitian quantum systems. Since Parity-time ($\mathcal{PT}%
$)-symmetric non-Hermitian quantum theory was put forward by Bender and
Boettcher\cite{Bender98,Bender02,Bender07} in the 1990s, it attracts massive
researches in different fields in recent years. Many platforms were proposed
to realize $\mathcal{PT}$-symmetric non-Hermitian symmetric quantum mechanics,
such as optical systems \cite{R10,Hang13,Peng16,Zhang16},
electronics\cite{Bender13b,Assawaworrarit17,Choi18},
microwaves\cite{Bittner12}, acoustics\cite{Zhu14,Popa14,Fleury15} and
single-spin system\cite{Wu19}. Some applications associated with
$\mathcal{PT}$-symmetric system have been explored including unidirectional
transport\cite{Feng13,Peng14} and single-mode lasers\cite{Feng14,Hodaei14}.
Also $\mathcal{PT}$-symmetric non-Hermitian quantum spin models have been
designed\cite{Korff07,Korff08,Giorgi10,Song13,Ghatak19}, which can well
capture the essence of many discrete models. For a (or anti) $\mathcal{PT}%
$-symmetric non-Hermitian Hamiltonian\cite{Ge13} which is invariant under the
combined action of the $\mathcal{P}$ and $\mathcal{T}$ operations, (or anti)
$\mathcal{PT}$ spontaneous symmetry breaking (SSB) occurs with adjustable
parameters accompanied by a real-to-complex spectral phase transition. The
critical point is called, exceptional point (EP), at which two or more
eigenvalues, and their corresponding eigenvectors, simultaneously coalesce.

As another typical quantum spin non-Hermitian models, rotation-time
($\mathcal{RT}$)-symmetric non-Hermitian spin models have been
studied\cite{Song13}. $\mathcal{RT}$-invariant non-Hermitian Hamiltonians for
quantum spin models as a class of pseudo-Hermitian Hamiltonians behave
homologous to (or anti) $\mathcal{PT}$-invariant non-Hermitian Hamiltonians.
It is naturally conceivable that the spin rotation operator $\mathcal{R}$
might replace the parity operator $\mathcal{P}$ to construct a quantum spin
model of non-Hermitian quantum systems. Therefore, $\mathcal{RT}$-invariant
non-Hermitian Hamiltonians for quantum spin models can be regarded as a
generalization of $\mathcal{PT}$-invariant non-Hermitian Hamiltonians. In
addition, the Hamiltonians with $\mathcal{RT}$ $\mathcal{(PT)}$-symmetric
complex magnetic fields between spins in quantum spin models
\cite{Giorgi10,Song13,Castro09,Deguchi09} are also non-Hermitian quantum
\textbf{systems} predicted to contain those properties as above mentioned.

In this paper, we investigate a one dimensional (1D) spin
model\cite{Giorgi10,Song13,Castro09,Deguchi09,Song16} with $\mathcal{RT}%
$-invariant non-Hermitian term and focus on how Non-Hermitian terms
affect\textbf{ }the degenerate ground states of this model. we find that the
effective Hamiltonian for the degenerate ground states has (or anti)
$\mathcal{PT}$-symmetry and there exists (or anti) $\mathcal{PT}$ SSB for the
two degenerate ground states with tunable parameters in external fields at
EPs. At EPs, the \textbf{original} real eigenvalues turn into the complex ones
and their associated eigenvectors simultaneously merge into one. By using high
order perturbation method\cite{kou09a,kou09b,kou09c} to manipulate the quantum
tunneling effect between two degenerate ground states, we detect that the
energy splitting of two degenerate ground states comes from quantum tunneling
in transverse filed and Zeeman splitting in longitudinal field, which is
precisely consistent with the numerical calculations of the extent of the
energy splitting of two degenerate ground states.

\section{The non-Hermitian Ising model with $\mathcal{RT}$-symmetry}

Let us begin with the 1D non-Hermitian Hamiltonian with $\mathcal{RT}%
$-symmetry for quantum spin model given by%
\begin{equation}
\hat{H}_{\mathcal{RT}}=\hat{H}_{0}+\hat{H}^{\prime}, \label{eq.1}%
\end{equation}
where $\hat{H}_{0}=-g\sum_{i}\tau_{i}^{z}\tau_{i+1}^{z}$ is 1D Ising model
without external field and $\hat{H}^{\prime}=\sum_{i}(\alpha\tau_{i}^{x}%
+\beta\tau_{i}^{z})$ represents the external field term. $\tau_{i}^{x,y,z}$
are Pauli matrices at sites $i$ and satisfy the periodic boundary condition
$\tau_{i}^{x,y,z}=\tau_{i+N}^{x,y,z}$.$\ \alpha$ and $\beta$ are the strength
of the transverse and longitudinal fields, respectively. As the perturbation
term, $\left\vert \alpha\right\vert ,$ $\left\vert \beta\right\vert $ are much
smaller than $g$ normally. In this paper, we focus on the case of
$g\gg\left\vert \alpha\right\vert $, $g\gg\left\vert \beta\right\vert $, and
the coupling parameter $g$ is set to be unit, $g\equiv1$. In particular,
$\alpha$ or $\beta$ can be real or imaginary tunable parameter.

For a non-Hermitian Hamiltonian with $\mathcal{RT}$-symmetry, we have $\left[
\mathcal{R},\hat{H}_{\mathcal{RT}}\right]  \neq0$ and $\left[  \mathcal{T}%
,\hat{H}_{\mathcal{RT}}\right]  \neq0$, but $\left[  \mathcal{RT},\hat
{H}_{\mathcal{RT}}\right]  =0.$ Here the time reversal operator $\mathcal{T}$
is defined as $\mathcal{T}i\mathcal{T=-}i$ and the spin rotation operator
$\mathcal{R}$ is defined by rotating each spin by $\pi$ about the $a$-axis
\[
\mathcal{R}=\mathcal{R}^{a}\mathcal{(}\pi\mathcal{)}=\Pi_{j=1}^{N}(i\tau
_{j}^{a}),
\]
where $N$ is the number of lattice sites and $a$ denotes $x$, $y$, $z$ directions.

In this paper, will study the following three cases under periodic boundary condition:

\begin{enumerate}
\item The non-Hermitian Ising model with $\mathcal{R}^{x}\mathcal{T}%
$-symmetry: for the case of $\mathcal{R}=\mathcal{R}^{x}\mathcal{(}%
\pi\mathcal{)}=\Pi_{j=1}^{N}(i\tau_{j}^{x})$, the spin rotation operator
$\mathcal{R}^{x}\mathcal{(}\pi\mathcal{)}$ turns into parity operation and the
$\mathcal{R}^{x}\mathcal{T}$-symmetry turns into traditional $\mathcal{PT}%
$-symmetry. Now, $\alpha$ is real and $\beta$ is imaginary, respectively. The
Hamiltonian becomes
\begin{equation}
\hat{H}_{\mathcal{R}^{x}\mathcal{T}}=-g\sum_{i}\tau_{i}^{z}\tau_{i+1}^{z}%
+\sum_{i}(\pm\left\vert \alpha\right\vert \tau_{i}^{x}\pm i\left\vert
\beta\right\vert \tau_{i}^{z}).
\end{equation}
This model has been applied to investigate a lattice version of the Yang--Lee
model\cite{Yang52,Lee52,Castro09,Deguchi09,Cardy85,Gehlen91a,Gehlen91b}.

\item The non-Hermitian Ising model with $\mathcal{R}^{z}\mathcal{T}%
$-symmetry: for the case of $\mathcal{R}=\mathcal{R}^{z}\mathcal{(}%
\pi\mathcal{)}=\Pi_{j=1}^{N}(i\tau_{j}^{z})$, the spin rotation operator
$\mathcal{R}^{z}\mathcal{(}\pi\mathcal{)}$ turns into anti-parity operation
and the $\mathcal{RT}$-symmetry turns into traditional anti-$\mathcal{PT}%
$-symmetry. Now, $\alpha$ is imaginary and $\beta$ is real, respectively. The
Hamiltonian becomes
\begin{equation}
\hat{H}_{\mathcal{R}^{z}\mathcal{T}}=-g\sum_{i}\tau_{i}^{z}\tau_{i+1}^{z}%
+\sum_{i}(\pm i\left\vert \alpha\right\vert \tau_{i}^{x}\pm\left\vert
\beta\right\vert \tau_{i}^{z});\nonumber
\end{equation}

\item The non-Hermitian Ising model with $\mathcal{R}^{y}\mathcal{T}%
$-symmetry: for the case of $\mathcal{R}=\mathcal{R}^{y}\mathcal{(}%
\pi\mathcal{)}=\Pi_{j=1}^{N}(i\tau_{j}^{y})=i\mathcal{R}^{x}\mathcal{(}%
\pi\mathcal{)R}^{z}\mathcal{(}\pi\mathcal{)}$, the spin rotation operator
$\mathcal{R}^{y}\mathcal{(}\pi\mathcal{)}$ turns into "parity" operation. Now,
$\alpha$ and $\beta$ are all imaginary. The Hamiltonian becomes
\begin{equation}
\hat{H}_{\mathcal{R}^{y}\mathcal{T}}=-g\sum_{i}\tau_{i}^{z}\tau_{i+1}^{z}%
+\sum_{i}(\pm i\left\vert \alpha\right\vert \tau_{i}^{x}\pm i\left\vert
\beta\right\vert \tau_{i}^{z}).
\end{equation}

\end{enumerate}

\section{Effective Hamiltonian for the (quasi) degenerate ground states}

In the limit of $\alpha,\beta\rightarrow0$, the Hamiltonian $\hat
{H}_{\mathcal{RT}}$ will reduce to 1D Ising model $\hat{H}_{0}$ without
external filed. Now the ground states have two-fold degeneracy and becomes
Ferromagnetic (FM) states, i.e., $\left\vert \uparrow\uparrow\cdots
\uparrow\uparrow\right\rangle $ and $\left\vert \downarrow\downarrow
\cdots\downarrow\downarrow\right\rangle $. Due to the existence of the
external fields, the energies for the FM degenerate ground states split. For
the case of $\alpha\neq0,$ $\beta=0$, the energy splitting for the degenerate
ground states comes from quantum tunneling effect. The quantum tunneling
process corresponds to the creation of a pair of virtual domain wall with one
of them traversing around the chain and annihilating with the other one
finally. Under the periodic boundary condition for the Ising chain, the
dominant tunneling process is a single virtual domain wall moving from one
side to the other with a reduced tunneling splitting obtained by high-order
perturbation approach. On the other hand, for the case of $\alpha=0,$
$\beta\neq0$, the energy splitting for the degenerate ground states comes from
Zeeman effect that leads to an energy splitting proportional to external field.

Now, we discuss the energy splitting from imaginary (or real) transverse
external field, i.e., the term $\sum_{i}\alpha\tau_{i}^{x}.$

According to high-order perturbation approach, after adding the transverse
field $\tau_{i}^{x}$ to the Ising chain, we are able to generate two domain
walls and drive one of them hopping around the chain by considering high-order
perturbation terms.

We denote the two ground states by $\left\vert \uparrow\uparrow\cdots
\uparrow\uparrow\right\rangle $ and $\left\vert \downarrow\downarrow
\cdots\downarrow\downarrow\right\rangle $. Considering this tunneling process,
we obtain the perturbation energy as we may obtain the energy shift $\Delta$
as
\begin{equation}
\Delta=\langle\uparrow\uparrow\cdots\uparrow\uparrow\mid\hat{H}^{\prime}%
(\frac{1}{E_{0}-\hat{H}_{0}}\hat{H}^{\prime})^{N-1}\left\vert \downarrow
\downarrow\cdots\downarrow\downarrow\right\rangle ,
\end{equation}
where $E_{0}$ is the ground state energy. According to $\hat{H}_{0}\left\vert
\downarrow\downarrow\cdots\uparrow\cdots\downarrow\downarrow\right\rangle
=(E_{0}+4g)\left\vert \downarrow\downarrow\cdots\uparrow\cdots\downarrow
\downarrow\right\rangle ,$ the energy for the excited state $\left\vert
\downarrow\downarrow\cdots\uparrow\cdots\downarrow\downarrow\right\rangle $ of
the two domain walls at the sites $i$ and $i-1$ is $E_{0}+4g$. Then we have
\begin{equation}
(\frac{1}{E_{0}-\hat{H}_{0}}\hat{H}^{\prime})\left\vert \downarrow
\downarrow\cdots\downarrow\downarrow\right\rangle =N(\frac{\alpha}%
{-4g})\left\vert \downarrow\downarrow\cdots\uparrow\cdots\downarrow
\downarrow\right\rangle .
\end{equation}
After the domain wall moves step by step around the chain and annihilates with
the other, the ground state changes from $\left\vert \downarrow\downarrow
\cdots\downarrow\downarrow\right\rangle $ to $\mid\uparrow\uparrow
\cdots\uparrow\uparrow\rangle$ (or from $\mid\uparrow\uparrow\cdots
\uparrow\uparrow\rangle$ to $\left\vert \downarrow\downarrow\cdots
\downarrow\downarrow\right\rangle $). As a result, the corresponding energy
splitting is obtained as
\begin{equation}
\Delta E=2\Delta=2\times N\frac{(\alpha)^{N}}{(-4g)^{N-1}}.
\end{equation}

Next, we consider the energy splitting from imaginary (or real) longitudinal
field $\sum_{i}\beta\tau_{i}^{z}$. When we apply the longitudinal external
field ($\beta\neq0$), the corresponding eigenstates of the Hamiltonian are
just $\left\vert \uparrow\uparrow\cdots\uparrow\uparrow\right\rangle $ and
$\left\vert \downarrow\downarrow\cdots\downarrow\downarrow\right\rangle $ with
energy shifting. The energy difference between the states $\left\vert
\uparrow\uparrow\cdots\uparrow\uparrow\right\rangle $ and $\left\vert
\downarrow\downarrow\cdots\downarrow\downarrow\right\rangle $ is obtained as
\begin{equation}
\Delta E=2\varepsilon=2N\beta.
\end{equation}

Finally, an effective pseudo-spin Hamiltonian for the (quasi) degenerate
ground states is obtained as
\begin{align}
\mathcal{\hat{H}}_{\mathrm{eff}} &  =\frac{\Delta}{2}(\left\vert
\uparrow\uparrow\cdots\uparrow\uparrow\right\rangle \langle\downarrow
\downarrow\cdots\downarrow\downarrow\mid\\
+\left\vert \downarrow\downarrow\cdots\downarrow\downarrow\right\rangle
\langle &  \uparrow\uparrow\cdots\uparrow\uparrow\mid)+\frac{\varepsilon}%
{2}(\left\vert \uparrow\uparrow\cdots\uparrow\uparrow\right\rangle
\langle\uparrow\uparrow\cdots\uparrow\uparrow\mid\nonumber\\
-\left\vert \downarrow\downarrow\cdots\downarrow\downarrow\right\rangle
\langle &  \downarrow\downarrow\cdots\downarrow\downarrow\mid)\nonumber\\
&  =\frac{\Delta}{2}\sigma^{x}+\frac{\varepsilon}{2}\sigma^{z},\nonumber
\end{align}
where $\Delta=N\frac{(\alpha)^{N}}{(-4g)^{N-1}}$ and $\varepsilon=N\beta$.
\textbf{In thermodynamic limit, }$\Delta\rightarrow0$\textbf{, we have }%
\begin{equation}
\mathcal{\hat{H}}_{\mathrm{eff}}\simeq\frac{\varepsilon}{2}\sigma^{z}\nonumber
\end{equation}
\textbf{t}hat is Hermitian when\textbf{ }$\operatorname{Im}\varepsilon
=0$\textbf{ }or anti-Hermitian when\textbf{ }$\operatorname{Re}\varepsilon
=0$\textbf{.}

By diagonalizing the effective Hamiltonian $\mathcal{\hat{H}}_{\mathrm{eff}}$,
we get eigenvalues and eigenvectors as
\[
E_{\pm}=\pm\sqrt{(\frac{\Delta}{2})^{2}+(\frac{\varepsilon}{2})^{2}}.
\]
The total energy splitting becomes
\begin{equation}
\Delta E=\left\vert E_{+}-E_{-}\right\vert =\sqrt{\Delta^{2}+\varepsilon^{2}}.
\end{equation}

\section{Spontaneous $\mathcal{PT}$-symmetry breaking for degenerate ground
states of the non-Hermitian Ising model with $\mathcal{R}^{x}\mathcal{T}%
$-symmetry}

Firstly, we consider the case of real $\alpha$ (transverse field) and
imaginary $\beta$ (Zeeman filed). The Hamiltonian of the\textit{\ }%
non-Hermitian Ising model becomes
\begin{equation}
\hat{H}_{\mathcal{R}^{x}\mathcal{T}}=-g\sum_{i}\tau_{i}^{z}\tau_{i+1}^{z}%
+\sum_{i}(\pm\left\vert \alpha\right\vert \tau_{i}^{x}\pm i\left\vert
\beta\right\vert \tau_{i}^{z}).
\end{equation}
The Hamiltonian has $\mathcal{R}^{x}\mathcal{T}$-symmetry, i.e, $\left[
\mathcal{R}^{x},\hat{H}_{\mathcal{R}^{x}\mathcal{T}}\right]  \neq0$ and
$\left[  \mathcal{T},\hat{H}_{\mathcal{R}^{x}\mathcal{T}}\right]  \neq0,$ but
$\left[  \mathcal{R}^{x}\mathcal{T},\text{ }\hat{H}_{\mathcal{R}%
^{x}\mathcal{T}}\right]  =0.$\ From above discussion, the effective
Hamiltonian for the (quasi) degenerate ground states is obtained as
\begin{equation}
\mathcal{\hat{H}}_{\mathrm{eff}}=\frac{\Delta}{2}\sigma^{x}+\frac{\varepsilon
}{2}\sigma^{z}=N\frac{(\left\vert \alpha\right\vert )^{N}}{(-4g)^{N-1}}%
\sigma^{x}\pm iN\left\vert \beta\right\vert \sigma^{z}.\nonumber
\end{equation}
There exists effective $\mathcal{PT}$-symmetry for the effective Hamiltonian
for the (quasi) degenerate ground states, i.e., $\left[  \mathcal{P}%
_{\mathrm{eff}},\mathcal{\hat{H}}_{\mathrm{eff}}\right]  \neq0$ and $\left[
\mathcal{T},\mathcal{\hat{H}}_{\mathrm{eff}}\right]  \neq0,$ but $\left[
\mathcal{P}_{\mathrm{eff}}\mathcal{T},\text{ }\mathcal{\hat{H}}_{\mathrm{eff}%
}\right]  =0$ where $\mathcal{P}_{\mathrm{eff}}=\sigma^{x}$.

For $\left\vert \Delta\right\vert $ $>\left\vert \varepsilon\right\vert $,
both eigenvalues $E_{\pm}$ are real, indicating the system in a phase with
effective $\mathcal{PT}$ symmetry. In the region of unbroken effective
$\mathcal{PT}$ symmetry, the eigenvectors are eigenstates of the symmetry
operator, i.e., $\mathcal{P}_{\mathrm{eff}}\mathcal{T}\psi_{\pm}=\psi_{\pm}$;
For $\left\vert \Delta\right\vert $ $<\left\vert \varepsilon\right\vert $,
both eigenvalues become imaginary. In this region $\theta$ is complex and
$\psi_{\pm}$ no longer possess the same symmetry as $\mathcal{\hat{H}%
}_{\mathrm{eff}}$. A spontaneous (effective) $\mathcal{PT}$-symmetry-breaking
transition occurs at the exceptional points $\left\vert \Delta\right\vert $
$=\left\vert \varepsilon\right\vert $ that indicates a relationship
\begin{equation}
\frac{\left\vert \alpha\right\vert ^{N}}{(4g)^{N-1}}=\left\vert \beta
\right\vert .
\end{equation}
As a result, the two degenerate ground states merge into one at EP.

\begin{figure}[ptb]
\includegraphics[clip,width=0.48\textwidth]{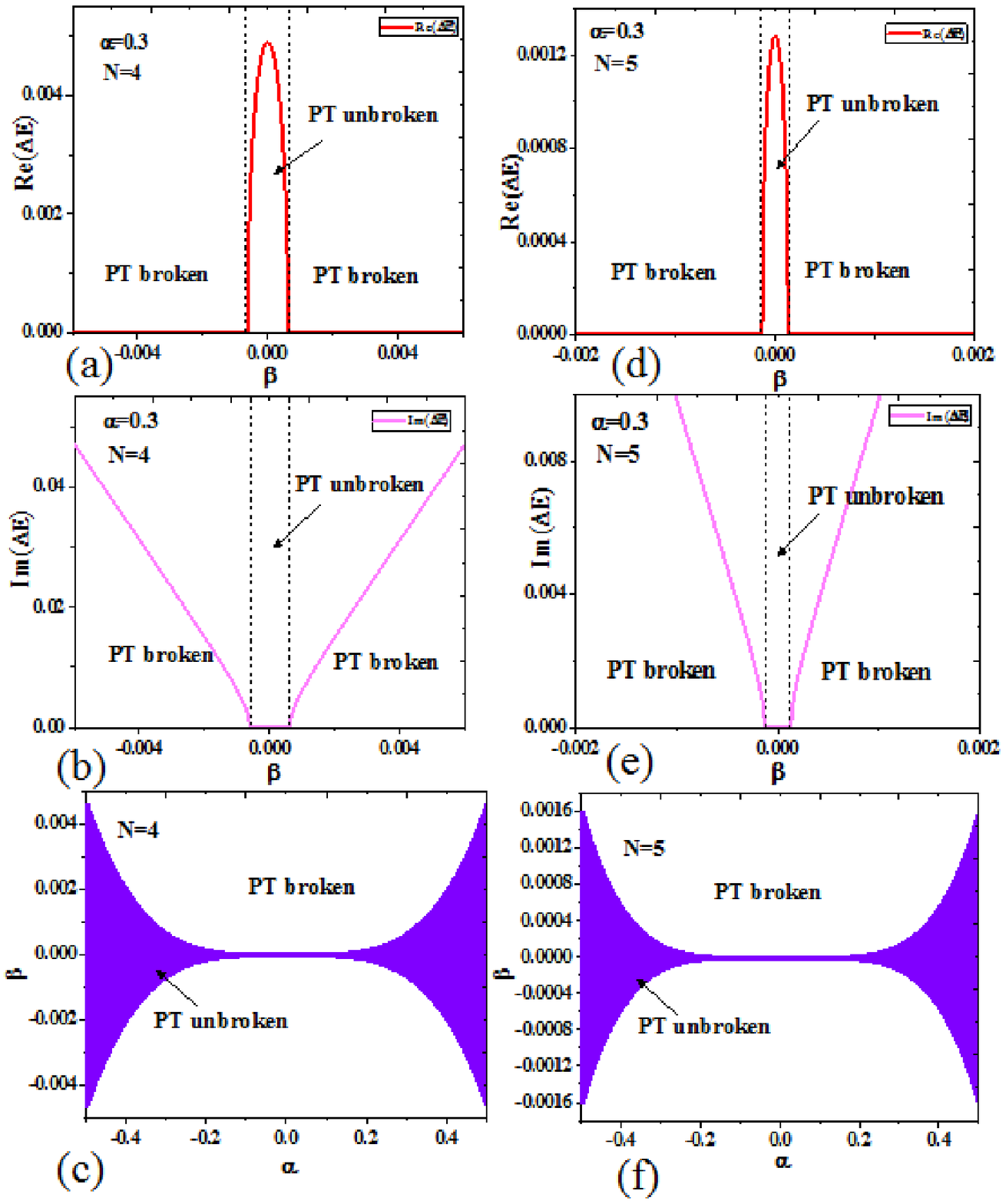}\caption{(Color online)
The effective $\mathcal{PT}$ phase transition under $\mathcal{R}%
^{x}\mathcal{T}$-symmetry. (a) and (b) The real and imaginary parts of energy
splitting between the two degenerate ground states, $\operatorname{Re}%
\left\vert \delta E\right\vert $ and $\operatorname{Im}\left\vert \delta
E\right\vert $ as a function of $\beta$ for the case of $\left\vert
\alpha\right\vert =0.3$ for even lattice $N=4$. (c) The phase diagram for two
degenerate ground states, the phase boundary is composed of exceptional points
for $N=4$. (d) and (e) The real and imaginary parts of energy splitting for
odd lattice $N=5$. (f) The phase diagram for two degenerate ground states for
$N=5$. }%
\label{Fig.1}%
\end{figure}

In Fig.~\ref{Fig.1}, we illustrate the numerical results from the exact
diagonalization technique of the Ising model on even and odd lattices with
periodic boundary conditions.\ For case of $N=4$, Fig.~\ref{Fig.1}(a) and (b)
show the real part $\operatorname{Re}\left\vert \delta E\right\vert $ and the
imaginary part $\operatorname{Im}\left\vert \delta E\right\vert $ for the
ground states energy splitting of the non-Hermitian Ising model with
$\left\vert \alpha\right\vert =0.3$, respectively. Fig.~\ref{Fig.1}(c) shows
the global phase diagram of $\mathcal{PT}$-symmetry-breaking transition for
the two degenerate ground states, of which the phase boundary is composed of
exceptional points characterized by the relation $\left\vert \alpha\right\vert
^{4}=(4g)^{3}\left\vert \beta\right\vert $. For case of $N=5$,
Fig.~\ref{Fig.1}(d) and Fig.~\ref{Fig.1}(e) show the real part
$\operatorname{Re}\left\vert \delta E\right\vert $ and the imaginary part
$\operatorname{Im}\left\vert \delta E\right\vert $ for the ground states
energy splitting of the non-Hermitian Ising model with $\left\vert
\alpha\right\vert =0.3$, respectively. Fig.~\ref{Fig.1}(f) shows the global
phase diagram of $\mathcal{PT}$-symmetry-breaking transition for degenerate
ground states, the phase boundary are exceptional points characterized by the
relation $\left\vert \alpha\right\vert ^{5}=(4g)^{4}\left\vert \beta
\right\vert $. We can see the numerical results are consistent to our
theoretical prediction by this high-order perturbation approach.

Therefore, we found that, quantum properties for $\hat{H}_{\mathcal{R}%
^{x}\mathcal{T}}$ on a spin chain with even number of lattice sites and those
for $\hat{H}_{\mathcal{R}^{x}\mathcal{T}}$ on a spin chain with odd number of
lattice sites are similar and even a small imaginary external field applied
will drive $\mathcal{PT}$ phase transition by reason of the Pauli matrices
non-commutative relation. 

\section{Spontaneous anti-$\mathcal{PT}$-symmetry breaking for degenerate
ground states of the non-Hermitian Ising model with $\mathcal{R}%
^{z}\mathcal{T}$-symmetry}

Secondly, we consider the case of imaginary $\alpha$ (transverse field) and
real $\beta$ (Zeeman filed). The Hamiltonian for the non-Hermitian Ising model
becomes
\begin{equation}
\hat{H}_{\mathcal{R}^{z}\mathcal{T}}=-g\sum_{i}\tau_{i}^{z}\tau_{i+1}^{z}%
+\sum_{i}(\pm i\left\vert \alpha\right\vert \tau_{i}^{x}\pm\left\vert
\beta\right\vert \tau_{i}^{z}).
\end{equation}
The Hamiltonian has $\mathcal{R}^{z}\mathcal{T}$-symmetry, i.e, $\left[
\mathcal{R}^{z},\hat{H}_{\mathcal{R}^{z}\mathcal{T}}\right]  \neq0$ and
$\left[  \mathcal{T},\hat{H}_{\mathcal{R}^{z}\mathcal{T}}\right]  \neq0,$ but
$\left[  \mathcal{R}^{z}\mathcal{T},\text{ }\hat{H}_{\mathcal{R}%
^{z}\mathcal{T}}\right]  =0.$\ From above discussion, the effective
Hamiltonian for the (quasi) degenerate ground states is obtained as
\begin{equation}
\mathcal{\hat{H}}_{\mathrm{eff}}=\frac{\Delta}{2}\sigma^{x}+\frac{\varepsilon
}{2}\sigma^{z}=N\frac{(i\left\vert \alpha\right\vert )^{N}}{(-4g)^{N-1}}%
\sigma^{x}+N\left\vert \beta\right\vert \sigma^{z}.\nonumber
\end{equation}

We found that, quantum properties for $\hat{H}_{\mathcal{R}^{z}\mathcal{T}}$
on a spin chain with even number of lattice sites and those for $\hat
{H}_{\mathcal{R}^{z}\mathcal{T}}$ on a spin chain with odd number of lattice
sites are quite different.

\begin{figure}[ptb]
\includegraphics[clip,width=0.48\textwidth]{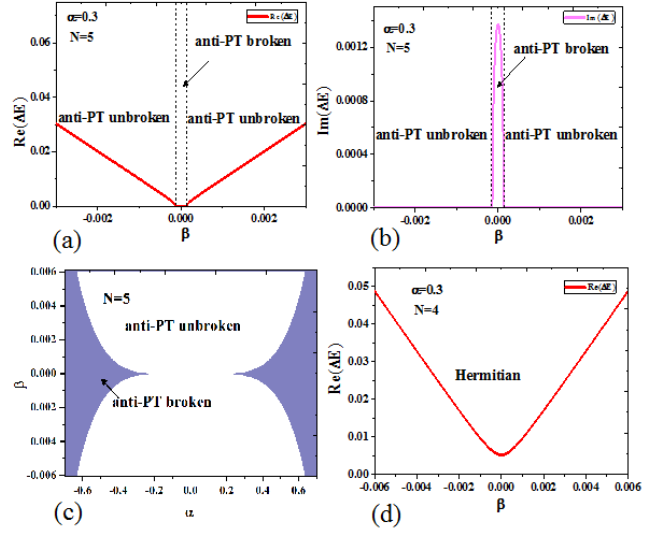}\caption{(Color online)
The effective anti-$\mathcal{PT}$ phase transition under $\mathcal{R}%
^{z}\mathcal{T}$-symmetry. (a) and (b) The real and imaginary parts of energy
splitting between the two degenerate ground states for the case of $\left\vert
\alpha\right\vert =0.3$ in case of odd lattice $N=5$. (c) The phase diagram
for the two degenerate ground states, the phase boundary is exceptional points
for $N=5$. (d) The energy splitting for even lattice $N=4$ under imaginary
transverse field.}%
\label{Fig.2}%
\end{figure}

In the case of a system on spin chain with even $N$ sites, the effective
Hamiltonian is Hermitian, i.e, $\mathcal{\hat{H}}_{\mathrm{eff}}%
=\mathcal{\hat{H}}_{\mathrm{eff}}^{\ast}.$ In Fig.~\ref{Fig.2}, we show the
numerical results from the exact diagonalization technique of the Ising model
with periodic boundary conditions. For $N=4$, the energy splitting $\Delta E$
is always real due to $(\frac{\Delta}{2})^{2}+(\frac{\varepsilon}{2})^{2}>0$,
the effective Hamiltonian is Hermitian, no $\mathcal{PT}$ phase transition happens.

Whereas, in the case of system on a spin chain with odd $N$ sites, the
effective Hamiltonian is non-Hermitian, i.e, $\mathcal{\hat{H}}_{\mathrm{eff}%
}\neq\mathcal{\hat{H}}_{\mathrm{eff}}^{\ast}$. There exists effective
anti-$\mathcal{PT}$-symmetry for the effective Hamiltonian for the (quasi)
degenerate ground states, i.e., $\left[  \mathcal{P}_{\mathrm{anti,eff}%
},\mathcal{\hat{H}}_{\mathrm{eff}}\right]  \neq0$ and $\left[  \mathcal{T}%
,\mathcal{\hat{H}}_{\mathrm{eff}}\right]  \neq0,$ but $\{ \mathcal{P}%
_{\mathrm{anti,eff}}\mathcal{T},$ $\mathcal{\hat{H}}_{\mathrm{eff}}\}=0$ where
$\mathcal{P}_{\mathrm{anti,eff}}=\sigma^{x}$. For $\left\vert \Delta
\right\vert $ $<\left\vert \varepsilon\right\vert $, we have effective
anti-$\mathcal{PT}$ symmetry. Now, both eigenvalues $E_{\pm}$ are real. In
this region of unbroken anti-$\mathcal{PT}$ symmetry, the eigenvectors are
eigenstates of the symmetry operator. For $\left\vert \Delta\right\vert $
$>\left\vert \varepsilon\right\vert $, effective anti-$\mathcal{PT}$ symmetry
is broken. A spontaneous (effective) anti-$\mathcal{PT}$-symmetry-breaking
transition occurs at the exceptional points $\left\vert \Delta\right\vert $
$=\left\vert \varepsilon\right\vert $ that indicates a power law relationship
\begin{equation}
\frac{\left\vert \alpha\right\vert ^{N}}{(4g)^{N-1}}=\left\vert \beta
\right\vert .
\end{equation}
As a result, at EP the two degenerate ground states also merge into one.

In Fig.~\ref{Fig.2}(d), we give the real parts of energy splitting for even
lattice $N=4$ , we derive the entirely real energy splitting but no imaginary
part of energy splitting due to its pure real effective Hermitian Hamiltonian.
Fig.~\ref{Fig.2}(a) and (b) illustrate $\operatorname{Re}\left\vert \delta
E\right\vert $ and $\operatorname{Im}\left\vert \delta E\right\vert $ of the
energy splitting for the two degenerate ground states, respectively. In
Fig.~\ref{Fig.2}(c), we show the global phase diagram of anti-$\mathcal{PT}%
$-symmetry-breaking transition for degenerate ground states of the
non-Hermitian Ising model with $\left\vert \alpha\right\vert =0.3$ on $N=5$
odd lattice. The phase boundary are exceptional points characterized by the
relation $\left\vert \alpha\right\vert ^{5}=(4g)^{4}\left\vert \beta
\right\vert $. We also find a very small imaginary external field applied will
drive anti-$\mathcal{PT}$ phase transition.

\section{Spontaneous $\mathcal{PT}$-symmetry breaking for degenerate ground
states of the non-Hermitian Ising model with $\mathcal{R}^{y}\mathcal{T}%
$-symmetry}

Thirdly, we consider the case of imaginary $\alpha$ (transverse field) and
imaginary $\beta$ (Zeeman filed). The Hamiltonian for the non-Hermitian Ising
model becomes
\begin{equation}
\hat{H}_{\mathcal{R}^{y}\mathcal{T}}=-g\sum_{i}\tau_{i}^{z}\tau_{i+1}^{z}%
+\sum_{i}(\pm i\left\vert \alpha\right\vert \tau_{i}^{x}\pm i\left\vert
\beta\right\vert \tau_{i}^{z}).
\end{equation}
The Hamiltonian has $\mathcal{R}^{y}\mathcal{T}$-symmetry, i.e, $\left[
\mathcal{R}^{y},\hat{H}_{\mathcal{R}^{y}\mathcal{T}}\right]  \neq0$ and
$\left[  \mathcal{T},\hat{H}_{\mathcal{R}^{y}\mathcal{T}}\right]  \neq0,$ but
$\left[  \mathcal{R}^{y}\mathcal{T},\text{ }\hat{H}_{\mathcal{R}%
^{y}\mathcal{T}}\right]  =0.$\ From above discussion, the effective
Hamiltonian for the (quasi) degenerate ground states is obtained as
\begin{equation}
\mathcal{\hat{H}}_{\mathrm{eff}}=\frac{\Delta}{2}\sigma^{x}+\frac{\varepsilon
}{2}\sigma^{z}=N\frac{(i\left\vert \alpha\right\vert )^{N}}{(-4g)^{N-1}}%
\sigma^{x}+iN\left\vert \beta\right\vert \sigma^{z}.\nonumber
\end{equation}

We found that quantum properties for $\hat{H}_{\mathcal{R}^{y}\mathcal{T}}$ on
a spin chain with even $N$ and those for $\hat{H}_{\mathcal{R}^{y}\mathcal{T}%
}$ on a spin chain with odd $N$ odd lattices are also quite different.

For the case of a system on spin chain with odd number of lattice sites, the
effective Hamiltonian is pure imaginary and non-Hermitian, i.e, $\mathcal{\hat
{H}}_{\mathrm{eff}}=-\mathcal{\hat{H}}_{\mathrm{eff}}^{\ast}$, resulting in no
$\mathcal{PT}$ phase transition.

For the case of system on a spin chain with even number of lattice sites, the
effective Hamiltonian is non-Hermitian, i.e, $\mathcal{\hat{H}}_{\mathrm{eff}%
}\neq\mathcal{\hat{H}}_{\mathrm{eff}}^{\ast}$. There exists effective
$\mathcal{PT}$-symmetry for the effective Hamiltonian for the (quasi)
degenerate ground states, i.e., $\left[  \mathcal{P}_{\mathrm{eff}%
},\mathcal{\hat{H}}_{\mathrm{eff}}\right]  \neq0$ and $\left[  \mathcal{T}%
,\mathcal{\hat{H}}_{\mathrm{eff}}\right]  \neq0,$ but $\left[  \mathcal{P}%
_{\mathrm{eff}}\mathcal{T},\text{ }\mathcal{\hat{H}}_{\mathrm{eff}}\right]
=0$ where $\mathcal{P}_{\mathrm{eff}}=\sigma^{x}$. For $\left\vert
\Delta\right\vert >\left\vert \varepsilon\right\vert $, we have effective
$\mathcal{PT}$ symmetry, both eigenvalues $E_{\pm}$ are real. In this region
of unbroken effective $\mathcal{PT}$ symmetry, the eigenvectors are
eigenstates of the symmetry operator, i.e., $\mathcal{P}_{\mathrm{eff}%
}\mathcal{T}\psi_{\pm}=\psi_{\pm}$. For $\left\vert \Delta\right\vert $
$<\left\vert \varepsilon\right\vert $, effective $\mathcal{PT}$ symmetry is
broken, both eigenvalues become imaginary and correspond to a gain and a loss
eigenstate. A spontaneous (effective) $\mathcal{PT}$-symmetry-breaking
transition occurs at the exceptional points $\left\vert \Delta\right\vert
=\left\vert \varepsilon\right\vert $ that indicates a relationship
$\frac{\left\vert \alpha\right\vert ^{N}}{(4g)^{N-1}}=\left\vert
\beta\right\vert .$ At EP the two degenerate ground states also merge into one.

\begin{figure}[t]
\includegraphics[clip,width=0.48\textwidth]{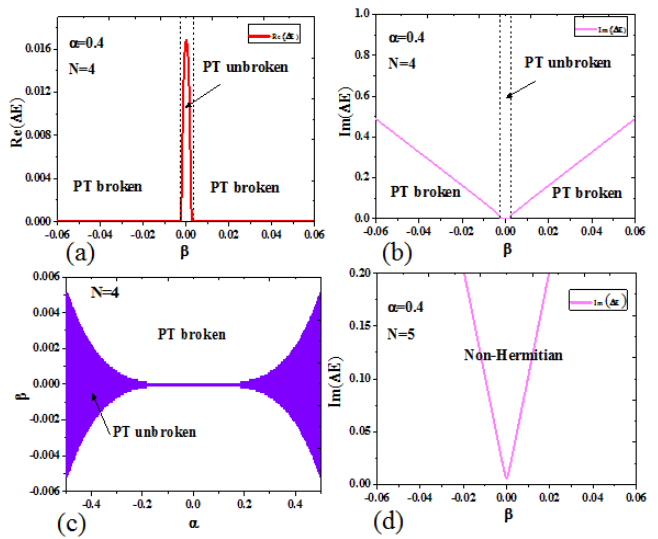}\caption{(Color online)
The effective $\mathcal{PT}$ phase transition under $\mathcal{R}%
^{y}\mathcal{T}$-symmetry. (a) and (b) The real and imaginary parts of energy
splitting between the two degenerate ground states for the case of $\left\vert
\alpha\right\vert =0.4$ with $N=4$. (c) The phase diagram for the two
degenerate ground states with $N=4$, the phase boundary is exceptional points.
(d) the imaginary parts of energy splitting between the two degenerate ground
states for the case of $\left\vert \alpha\right\vert =0.4$ with $N=5$.}%
\label{Fig.3}%
\end{figure}

In Fig.~\ref{Fig.3}(a) and (b) \ we depict the $\operatorname{Re}\left\vert
\delta E\right\vert $ and $\operatorname{Im}\left\vert \delta E\right\vert $
of energy splitting for the two degenerate ground states of the non-Hermitian
Ising model with $\left\vert \alpha\right\vert =0.4$ on even lattice $N=4$. In
Fig.~\ref{Fig.3}(c) we exhibit the global phase diagram of effective
anti-$\mathcal{PT}$-symmetry-breaking transition for degenerate ground states
for $N=4$ , the phase boundary is composed of exceptional points characterized
by the relation $\left\vert \alpha\right\vert ^{4}=(4g)^{3}\left\vert
\beta\right\vert $, Fig.~\ref{Fig.3}(d) shows the imaginary parts of energy
splitting for odd lattice $N=5$, we obtain the entirely imaginary energy
splitting but no real part of energy splitting due to its pure imaginary
non-Hermitian Hamiltonian.

\section{Conclusion and discussion}

\begin{table*}[th]
\caption{Non-Hermitian spin Ising model with different $\mathcal{RT}$
symmetries}%
\label{table.1}
\centering
\begin{tabular}
[c]{|l|l|}\hline\hline
$\hat{H}_{\mathcal{RT}}=\hat{H}_{0}+\sum_{i}(\alpha\tau_{i}^{x}\pm i\beta
\tau_{i}^{z})$ & $\mathcal{\hat{H}}_{\mathrm{eff}}=\frac{\Delta}{2}\sigma
^{x}+\frac{\varepsilon}{2}\sigma^{z}$\\\hline\hline
Real $\alpha$, imaginary $\beta$ ($\mathcal{R}^{x}\mathcal{T}$ symmetry) &
$\left\{
\begin{array}
[c]{l}%
\text{Even lattice: Real }\Delta\text{, imaginary }\varepsilon\text{
(}\mathcal{PT}\text{ symmetry)}\\
\text{Odd lattice: Real }\Delta\text{, imaginary }\varepsilon\text{
(}\mathcal{PT}\text{ symmetry)}%
\end{array}
\right.  $\\\hline\hline
Imaginary $\alpha$, real $\beta$ ($\mathcal{R}^{z}\mathcal{T}$ symmetry) &
$\left\{
\begin{array}
[c]{l}%
\text{Even lattice: real }\Delta\text{, real }\varepsilon\text{ }\\
\text{Odd lattice: imaginary }\Delta\text{, real }\varepsilon\text{
(anti-}\mathcal{PT}\text{ symmetry)}%
\end{array}
\right.  $\\\hline\hline
Imaginary $\alpha$, imaginary $\beta$ ($\mathcal{R}^{y}\mathcal{T}$
symmetry) & $\left\{
\begin{array}
[c]{l}%
\text{Even lattice: Real }\Delta\text{, imaginary }\varepsilon\text{
(}\mathcal{PT}\text{ symmetry)}\\
\text{Odd lattice: imaginary }\Delta\text{, imaginary }\varepsilon
\end{array}
\right.  $\\\hline\hline
\end{tabular}
\end{table*}

In this paper, we studied non-Hermitian spin Ising model with $\mathcal{RT}%
$-symmetry. To describe the low energy physics, we introduce an effective
pseudo-spin model, of which the (anti) $\mathcal{PT}$-symmetry emerges. In
particular, spontaneous (or anti) $\mathcal{PT}$-symmetry breaking may happen
in parameters space with tunable external field for special spin chains. As a
result, at EPs the two degenerate ground states always merge into one. These
results are consistent with those obtained from exact diagnalization numerical
technique. Also we find that $\mathcal{PT}$ phase transition is very robust
even in a tiny imaginary field although lattice sizes determine its phase
boundary. We can expand this conclusion to other spin systems with imaginary
external fields for studying their degenerate ground states properties due to
the Pauli matrices non-commutative relation.

We give Table. \ref{table.1} to show the main results -- the physics results
for non-Hermitian spin Ising model with different $\mathcal{RT}$-symmetries:
1) For the non-Hermitian Ising model with $\mathcal{R}^{x}\mathcal{T}%
$-symmetry and even/odd number of lattice sites, there exists effective
$\mathcal{PT}$-symmetry for the effective Hamiltonian for the (quasi)
degenerate ground states. Spontaneous $\mathcal{PT}$-symmetry breaking occurs;
2) For the non-Hermitian Ising model with $\mathcal{R}^{z}\mathcal{T}%
$-symmetry and even number of lattice sites, effective Hamiltonian for the
(quasi) degenerate ground states is Hermitian. However, for the non-Hermitian
Ising model with $\mathcal{R}^{z}\mathcal{T}$-symmetry and odd number of
lattice sites, effective Hamiltonian for the (quasi) degenerate ground states
is non-Hermitian with effective anti-$\mathcal{PT}$-symmetry. Spontaneous
anti-$\mathcal{PT}$-symmetry breaking occurs; 3) For the non-Hermitian Ising
model with $\mathcal{R}^{y}\mathcal{T}$-symmetry and odd number of lattice
sites, effective Hamiltonian for the (quasi) degenerate ground states is
non-Hermitian and pure imaginary. However, for the non-Hermitian Ising model
with $\mathcal{R}^{y}\mathcal{T}$-symmetry and even number of lattice sites,
effective Hamiltonian for the (quasi) degenerate ground states is
non-Hermitian with effective $\mathcal{PT}$-symmetry. Spontaneous
$\mathcal{PT}$-symmetry breaking occurs.

In the end, we address several relevant issues. The first is the relationship
between the non-Hermitian Ising model and the non-Hermitian systems with
topological bands. It was known that the quantum transverse Ising model is
equivalent to a topological fermionic model after Jordan-Wigner
transformation. However, for the 1D non-Hermitian Ising model with both
transverse and longitudinal fields, the Jordan-Wigner transformation doesn't
work and the quantum spin model cannot be mapped to a local (Hermitian or
non-Hermitian) fermionic model. Another relevant issue is the experimental
realization. It is still a challenge to experimentally investigate
non-Hermitian Hamiltonian related physics in quantum systems. A possible
approach is cold-atom experiments due to spontaneous decay\cite{Hang2013,
Lee2014,luo}. A possible application is to obtain a Schr\"{o}dinger cat state.
The basic idea is to\textbf{ }drive the two-level system of degenerate ground
states to EPs by adding external field and then remove it slowly. Due to the
non-Hermitian term, the degeneracy is reduced into non-Hermitian degeneracy at
the exceptional points. As a result, the two ground states merge into one
quantum "steady" state and a pure Schr\"{o}dinger cat state is
obtained.\textbf{ }

\acknowledgments This work is supported by NSFC Grant No. 11674026, 11974053.

\end{document}